\documentclass[a4paper,fleqn,usenatbib]{mnras}

\usepackage[T1]{fontenc}
\usepackage{ae,aecompl}

\usepackage{graphicx}	% Including figure files
\usepackage{amsmath}	% Advanced maths commands
\usepackage{amssymb}	% Extra maths symbols

\newcommand{\ci}{c_{\mathrm{i}}}
\newcommand{\Qzero}{Q_0}

\title[The expansion of spherical H{\sc\,ii} regions]{The classical D-type expansion of spherical H{\sc\,ii} regions}

\author[Williams et al.]
{\parbox{\textwidth}{Robin J. R. Williams$^{1}$\thanks{E-mail: \texttt{robin.williams@awe.co.uk}},
Thomas G. Bisbas$^{2,3}$, Thomas J. Haworth$^{4}$ \\ and Jonathan Mackey$^{5}$ 
}\vspace{0.4cm}\\
\parbox{\textwidth}{$^{1}$ AWE plc, Aldermaston, Reading RG7 4PR, UK\\
$^2$ Department of Astronomy, University of Florida, Gainesville, FL 32611, USA \\
$^3$ Max-Planck-Institut f\"ur Extraterrestrische Physik, Giessenbachstrasse 1, D-85748 Garching, Germany \\
$^{4}$ Astrophysics Group, Imperial College London, Blackett Laboratory, Prince Consort Road, London SW7 2AZ, UK\\
$^5$ Dublin Institute for Advanced Studies, Dunsink Observatory, Castleknock, Dublin 15, Ireland.
}}

\begin{document}
% These dates will be filled out by the publisher
\date{Accepted XXX. Received YYY; in original form ZZZ}

% Enter the current year, for the copyright statements etc.
\pubyear{2015}

% Don't change these lines

\label{firstpage}
\pagerange{\pageref{firstpage}--\pageref{lastpage}}
\maketitle

% Abstract of the paper
\begin{abstract}
  Recent numerical and analytic work has highlighted some shortcomings
  in our understanding of the dynamics of H{\sc\,ii} region expansion,
  especially at late times, when the H{\sc\,ii} region approaches
  pressure equilibrium with the ambient medium.  Here we reconsider
  the idealized case of a constant radiation source in a uniform and
  spherically symmetric ambient medium, with an isothermal equation of
  state.  A thick-shell solution is developed which captures the
  stalling of the ionization front and the decay of the leading shock
  to a weak compression wave as it escapes to large radii.  An
  acoustic approximation is introduced to capture the late-time damped
  oscillations of the H{\sc\,ii} region about the stagnation radius.
  Putting these together, a matched asymptotic equation is derived for
  the radius of the ionization front which accounts for both the
  inertia of the expanding shell and the finite temperature of the
  ambient medium.  The solution to this equation is shown to agree
  very well with the numerical solution at all times, and is superior
  to all previously published solutions.  The matched asymptotic
  solution can also accurately model the variation of H{\sc\,ii}
  region radius for a time-varying radiation source.

\end{abstract}

% Select between one and six entries from the list of approved keywords.
% Don't make up new ones.

%%%%%%%%%%%%%%%%%%%%%%%%%%%%%%%%%%%%%%%%%%%%%%%%%%

%%%%%%%%%%%%%%%%% BODY OF PAPER %%%%%%%%%%%%%%%%%%
\section{Introduction}

Massive stars are an important source of mass and energy to the
interstellar medium (ISM), through the radiation they emit, the strong
winds which they drive, and the supernova explosions by which they end
their lives.  Feedback from massive star formation can redistribute
matter, sculpting the ISM and potentially inhibiting or promoting star
formation \citep{2015MNRAS.454..238W,2015NewAR..68....1D}.  Supernovae
are the dominant feedback agent driving turbulent gas flows in
star-forming regions \citep{2004RvMP...76..125M}.  For a given
star-formation episode, however, ionizing radiation starts to be
emitted as soon as the first massive star reaches the zero-age main
sequence.  Heating by this ionizing radiation is the dominant feedback
process before supernovae commence for all but the most massive,
compact star clusters where the escape velocity is above
$20{\rm\,km\,s^{-1}}$
\citep{2007ASSP....1..103H,2014MNRAS.442..694D,2015ApJ...798...32N}.

The simplest model of the effect of photoionization on the ISM, a
spherically symmetric expansion of a bubble of ionized gas around
point source of ultraviolet photons with constant flux, is a classical
problem in ISM physics
\citep{1978ppim.book.....S,1997pism.book.....D,2006agna.book.....O}.
This system develops through a number of stages.  At first, each
photon emitted leads to an additional ionization, so the speed of
expansion of the ionization region is determined by the ionizing
luminosity.  Soon, the ionized material starts to recombine, and an
increasing fraction of the luminosity is taken up in maintaining the
level of ionization.  The expansion of the front slows as a result
both of geometric dilution of the ionizing radiation and the
increasing rate of recombinations in the ionized region.  Within the
ionized region, the pressure increases due to both the larger number
of free particles and the higher material temperature.  As a result,
when the rate of expansion of the ionization front slows to
approximately twice the sound speed within the ionized region, a
strong shock is generated at its surface which moves ahead of the
ionization front into the surrounding neutral material.  As time
progresses, this shock weakens and its separation from the ionization
front increases.  The ionization front slows and eventually relaxes to
an equilibrium radius at which the pressure in the rarefied ionized
region balances that in the unshocked external medium.

While clearly an idealization of the expansion of an H{\sc\,ii} region
around a young massive star, this model captures many essential
features of its dynamics.  The analytic formulation of this problem
follows classic work on ionization fronts \citep{1954BAN....12..187K}
by taking the ionized gas to be isothermal with a temperature
$T\approx10^4$\,K and the neutral gas also isothermal, but with a far
lower temperature ($T\approx 10\mbox{--}100{\rm\,K}$).

Early analytic and numerical studies of a spherically symmetric
H{\sc\,ii} region are summarised by \citet{1969ARA&A...7...67M}, by
which time ground-breaking numerical simulations had been performed
\citep{1966ApJ...143..700L} but an analytic form for the expansion as
a function of time had not been obtained.  An analytic solution for
the spherical expansion of an H{\sc\,ii} region was presented by
\citet{1978ppim.book.....S} by assuming the equality of total pressure
on either side of the swept-up shell, and this became the standard
solution \citep[e.g.][]{1997pism.book.....D}.  It was, however,
missing an essential concept: the inertia of the swept-up shell
\citep{2006ApJ...646..240H}, as had already been treated by
\citet{1977ApJ...214..725E}, albeit for a planar rather than spherical
front.  Simulations by \citet{2012MNRAS.419L..39R} found that the
ionization front overshoots the equilibrium radius at late times and
relaxes back, a behaviour which the Spitzer and Hosokawa \& Inutsuka
solutions do not capture.

\citet{2015MNRAS.453.1324B} compared results for this problem for
multiple time-dependent flow dynamics codes, finding good agreement
between them.  This paper also compared these computational results
with a number of analytical and semi-analytical models for the
expansion of the H{\sc\,ii} region.  At early time, where the shocked
shell is thin, much better agreement was found with the model of
\citet{2006ApJ...646..240H} than that of \citet{1978ppim.book.....S}.
Although Raga's modification of the Spitzer solution
\citet{2012RMxAA..48..149R}, referred to below as Raga-II, relaxes to
the correct radius of the H{\sc\,ii} region as $t\rightarrow\infty$,
none of the available models provides a good description of the later
time development of the H{\sc\,ii} region as it comes into pressure
equilibrium with its environment.  The intention of the present paper
is to remedy this.

While the late time relaxation of an H{\sc\,ii} region is more subject
to details such as the changing ultraviolet radiance of the central
star and the density distribution of the interstellar medium
environment, the idealized problem which we consider is not without
practical relevance.  H{\sc\,ii} regions form an important component
of the interstellar medium in the disks of late type galaxies.  The
shells which they drive are a source for the turbulent motions within
this medium \citep[e.g.][]{2009ApJ...694L..26G, 2014MNRAS.445.1797M, 2016MNRAS.463.2864A}. 
The nature of their response to variations in the pressure of the interstellar 
medium will also have important effects, for example, on the nature of the response 
of the medium to supernova explosions \citep[e.g.][]{2013MNRAS.431.1337R} and density 
waves in galactic disks.  It is also very useful to have reference solutions which 
capture the full evolution of an H{\sc\,ii} region for the purposes of validating 
computer algorithms.

In this paper, we first discuss the dimensionless parameters which
characterize the problem.  We then present our reference numerical
calculation, before developing a range of analytic and semi-analytic
treatments for the structure of the evolving H{\sc\,ii} region, which
we compare to the numerical solution.  The most detailed of these is
compared to additional numerical calculation where the ionization
source is changed during the evolution.  Finally, we summarize our
results, including the potential for their application as a sub-grid
model in larger scale simulations of the interstellar medium.

\section{Controlling parameters}

To define how wide a parameter space must be covered by an analytic
treatment for the problem of expansion of an H{\sc\,ii} region, it is
helpful to determine the dimensionless parameters which control the
process, as this will determine whether cases can be scaled to a
single common solution.

We consider the dimensional controlling parameters for the expansion
of a dust-free H{\sc\,ii} region in an isothermal environment free of
magnetic fields.  These are the particle density of the environment,
$n_0$, the source rate of ionizing photons $\Qzero$, and the hot and
cold material sound speeds, $\ci$ and $c_0$.  We use the sound speeds
and particle density to capture the dependency of the system on
molecular mass.  The relevant atomic physics is captured by the
ionization cross-section $a$ and the case B recombination coefficient
$\alpha_{\rm B}$, into excited states only.  Applying case B implies
that the H{\sc\,ii} region is assumed to be very optically thick to
Lyman continuum photons, and hence that the photons emitted by direct
recombinations to the $n=1$ level will ionize some neighbouring atom,
leading to no net recombination \citep[known as the `on-the-spot'
  approximation,][]{2006agna.book.....O}.  \citet{2009MNRAS.400..263W}
explicitly treat the radiation transfer of the diffuse Lyman
continuum, and confirm that this will typically have a negligible
effect on the structure of the H{\sc\,ii} region compared to the case
B approximation.

With six parameters, we are looking for four dimensionless parameters
for the problem.  The forms we will consider for these parameters are
\begin{equation}
  {\ci\over c_0}; \quad {\Qzero\over n\alpha_{\rm B}};
  \quad U\sim{n^{1/3}\Qzero^{1/3}\alpha_{\rm B}^{2/3}\over \ci};
  \quad {a \ci\over\alpha_{\rm B}}.
\end{equation}
The first two of these parameters have obvious interpretations: the
ratio of sound speeds determines the initial over-pressure (and
eventual under-density) of the H{\sc\,ii} region, and the second
parameter is the total number of particles in the initial nebula.

The third parameter $U$ is proportional to the ratio of the ionizing
photon flux to the thermal particle flux at the initial
\citep{1939ApJ....89..526S} radius, $R_{\rm St}$, if absorption were
neglected.  $R_{\rm St}$ satisfies
\begin{equation}
  \Qzero = {4\pi\over 3} \alpha_{\rm B} n^2 R_{\rm St}^3 \;.
\label{eqn:Rstro}
\end{equation}
Equivalently, $U$ is the ratio of the sound-crossing time of the
nebula to the characteristic time for ions to recombine.  It may also
be related to the column density through the ionized nebula, $N$, by
\begin{equation}
  U \sim {\alpha_{\rm B} N\over \ci}.
\end{equation}
This is a typical ionization parameter.

The fourth parameter is dependent only on atomic physics, so will not
vary substantially between observed nebulae.  The product of the third
and fourth parameters is independent of $\ci$, and determines the
ratio of the Str\"omgren radius 
to the ionization front thickness, $\sim naR_{\rm St}$
as well as the mean level of ionization within the nebula, since
locally
\begin{equation}
  {x^2\over 1-x} = {a J_{\rm i}\over n\alpha_{\rm B}},
\end{equation}
and $\left\langle J_{\rm i}\right\rangle \sim \Qzero/4 \pi R_{\rm St}^2$:
ionization fronts are geometrically thin for the same reason that
their cores are almost fully ionized, and what matters most for
dynamical studies is that both these statements are true.

Typical values of $\Qzero\simeq 10^{49}{\rm\,s^{-1}}$, $\alpha_{\rm B}
\simeq 2\times10^{-13}{\rm\,cm^3\,s^{-1}}$,
$a\simeq6\times10^{-18}{\rm\,cm^{-2}}$, $\ci = 10{\rm\,km\,s^{-1}}$,
$c_0 = 0.3{\rm\,km\,s^{-1}}$ give
\begin{eqnarray}
  {\ci\over c_0} &\simeq& 30,\\
  {\Qzero\over n\alpha_{\rm B}} &\simeq& {5\times 10^{61}\over n}, \\
  {n^{1/3}\Qzero^{1/3}\alpha_{\rm B}^{2/3}\over \ci} &\simeq& 74 n^{1/3},\\
  {a \ci\over\alpha_{\rm B}} &\simeq& 30.
\end{eqnarray}
The second parameter is so overwhelmingly large that the behaviour of
nebulae will almost always be in the asymptotic regime which means
that the structure of the nebula can be well modelled as a continuum
fluid.  The first and fourth parameters are weakly dependent on the
circumstances of a particular nebula, but have a similar scale to the
third parameter.  Hence the structure and evolution of H{\sc\,ii}
regions is only to be significantly affected by the value of the third
dimensionless parameter.

The structure may pass through a number of regimes through the
lifetime of an individual nebula, as the value of this parameter
varies with respect to that of the first and fourth parameters.
However, these effects will generally be quite weak, so long as the
mean ionization within the region remains close to complete (and if
the radiation pressure due to the ionizing continuum is ignored).
Varying the ionization parameter will affect the spectrum of emission
line radiation more significantly than it does the gross hydrodynamics
of the expansion of the H{\sc\,ii} region.

\section{Numerical calculation}

We will use the Early Phase test of the {\sc StarBench} project
presented in \cite{2015MNRAS.453.1324B} to compare with a number of
analytical approximations that will be discussed in detail in
Section~\ref{s:analytic}.  We have run the calculation beyond the latest
time considered in the previous paper, to allow the asymptotic
behaviour to be studied.  A larger domain has to be considered to
ensure that the out-going expansion wave remains on the mesh.

For reference, in this problem the density of the neutral medium
(consisting of pure hydrogen) is taken to be $\rho_{\rm
  o}=5.21\times10^{-21}\,{\rm g}\,{\rm cm}^{-3}$. We consider a source
emitting ionizing photons at a constant rate of $\Qzero=10^{49}\,{\rm
  s}^{-1}$. The resulting two-phase media have temperatures $T_{\rm
  i}=10^4\,{\rm K}$ and $T_0=10^2\,{\rm K}$ for the ionized
($\mu=0.5$) and neutral ($\mu=1$) regimes respectively corresponding
to sound speeds of $\ci=12.85\,{\rm km/s}$ and $c_0=0.91\,{\rm
  km/s}$. The initial Str\"omgren radius is $R_{\rm St}=0.314\,{\rm
  pc}$ and we evolve for $t=200\,{\rm Myr}$, which allows for several
oscillations of the H{\sc ii} region about the stagnation radius of
$R_{\rm Stag}=(\ci/c_0)^{4/3}R_{\rm St}=10.75\,{\rm pc}$
\citep{2012RMxAA..48..149R}.

We use the results of numerical calculations performed by the {\sc
  Glide} code.  {\sc Glide} is a spherically-symmetric Lagrangian code
which participated in the {\sc StarBench} project, and is described in
\citet{2015MNRAS.453.1324B}.  While this code has not been widely
applied, its results agree well with other codes in the {\sc
  StarBench} comparison, and its algorithm minimizes the effects of
numerical mixing on the evolution of the H{\sc\,ii} region compared to
solving on a mesh fixed in space.

\begin{figure}
\begin{center}
\begin{tabular}{l}
\includegraphics[width=\linewidth]{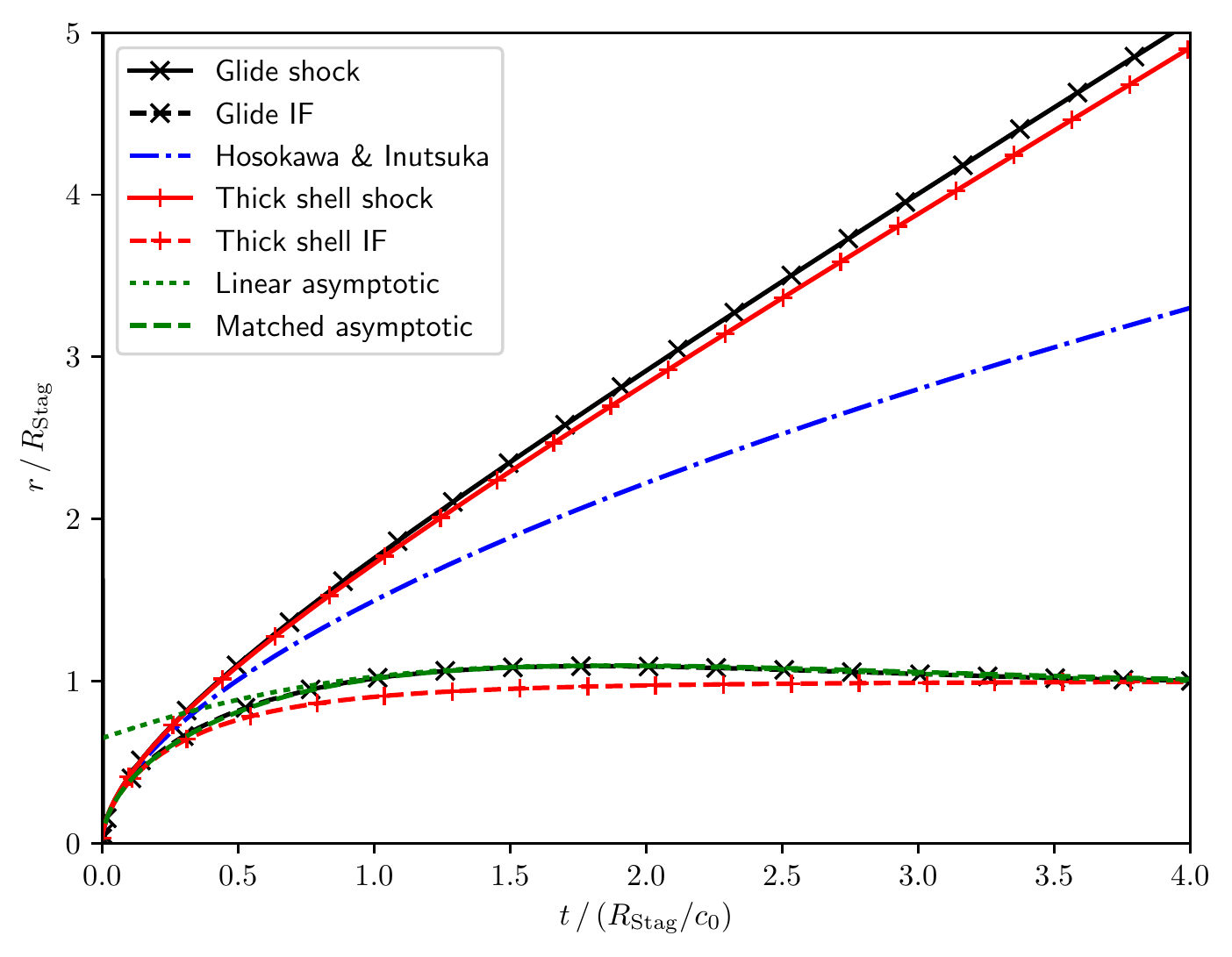} \\
\end{tabular}
\end{center}
\caption{Ionization and shock front positions from a fluid dynamical
  calculation with {\sc Glide}, compared to several approximations.
  The vertical axis is the radial distance as a fraction of $R_{\rm
    Stag}$, the horizontal axis the time as a function of the
  neutral-phase sound crossing time $R_{\rm Stag}/c_0$.}
\label{f:approx_if_shock}
\end{figure}

\begin{figure}
\begin{center}
\includegraphics[width=\linewidth]{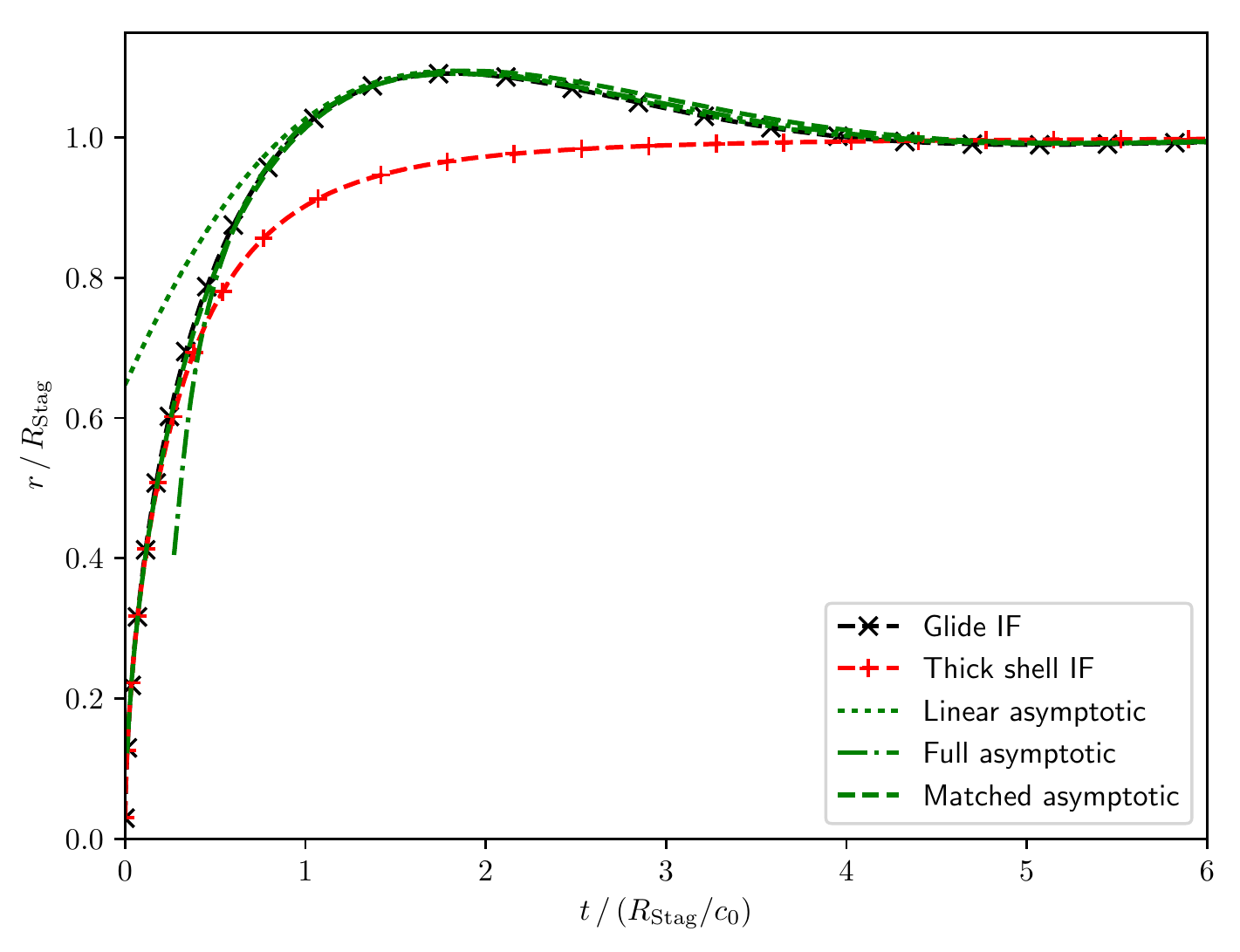}
\end{center}
\caption{As Figure~\protect\ref{f:approx_if_shock}, but scaled to
  emphasize early- to mid-time expansion of the ionization front.}
\label{f:approx_if_early}
\end{figure}

\begin{figure}
\begin{center}
\includegraphics[width=\linewidth]{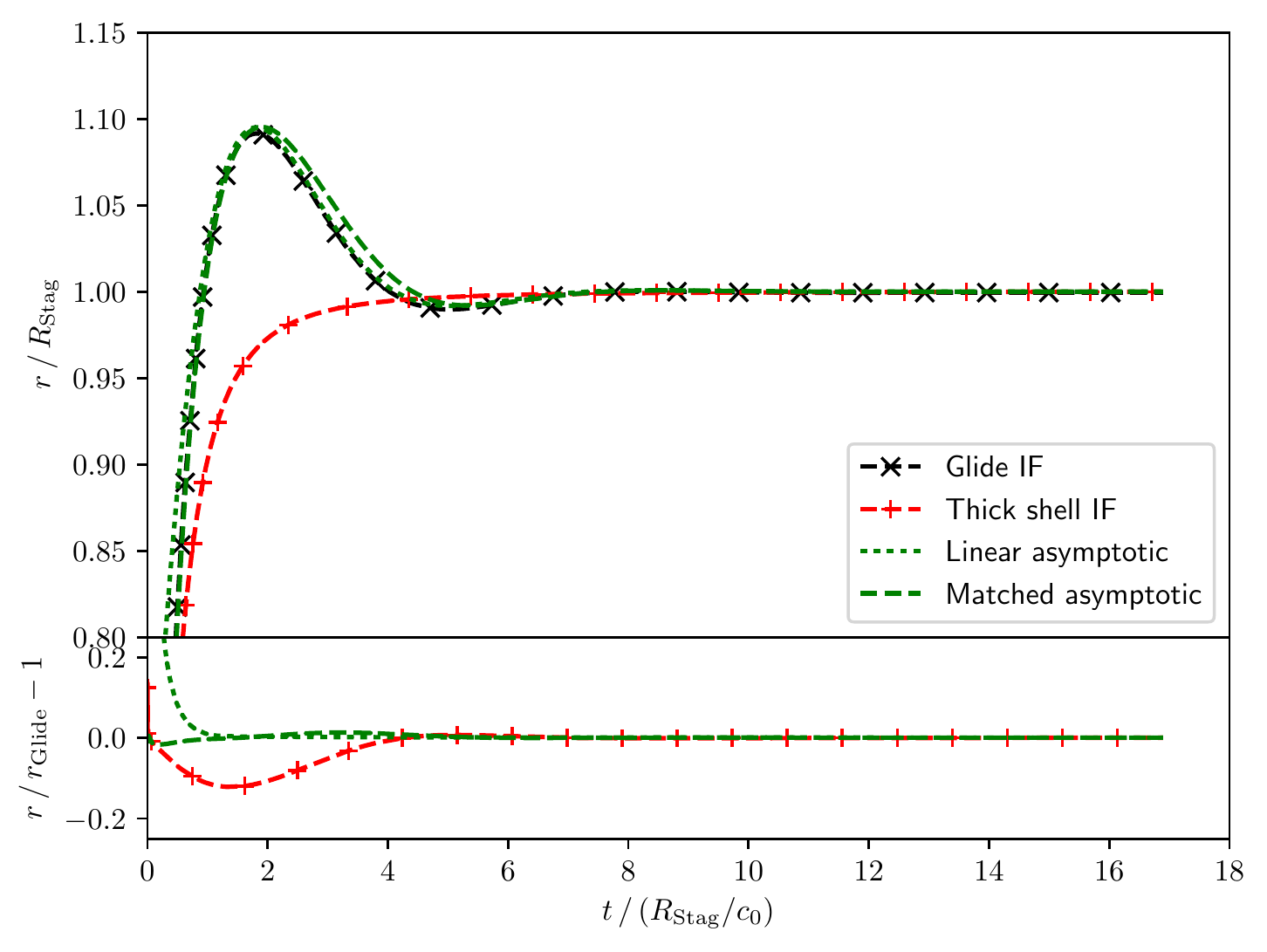}
\end{center}
\caption{As Figure~\protect\ref{f:approx_if_shock}, but scaled to
  emphasize late-time relaxation.  The lower panel shows the
  fractional residual between each of the approximations and the
  simulation result.}
\label{f:approx_if_late}
\end{figure}

To capture the different phases of evolution of this system, we show
three separate plots, Figures~\ref{f:approx_if_shock},
\ref{f:approx_if_early} and~\ref{f:approx_if_late}, which,
respectively, capture the thickening of the shocked shell, ionization
front overshoot and late time relaxation of the ionized bubble.  These
plots also show a number of analytical and semi-analytical models for
the position of the ionization and shock fronts, which we will discuss
in the following section.  Figure~\ref{f:approx_if_shock} shows how
the leading shock and ionization front initially move together.  As
the ionization front approaches the equilibrium expansion radius, both
fronts slow, and the shell between them becomes geometrically thick.
The speed of the shock front reduces asymptotically to the sound speed
in the neutral medium.  Figure~\ref{f:approx_if_early} shows that the
ionization front does expand beyond its equilibrium radius by a small
amount, as a result of the inertia of the material within the shell,
but subsequently relaxes back towards its equilibrium radius.
Figure~\ref{f:approx_if_late} shows that this relaxation is in fact a
strongly damped oscillation.

\section{Analytic models}
\label{s:analytic}

The development of an H{\sc\,ii} region passes through a number of
regimes with time, as different flow parameters become dominant.  In
this paper, we concentrate on the expansion of the region once the
ionization front has reached the weak D-type phase, where the shocked
neutral material ahead of the front is close to pressure equilibrium
with the ionized material around the central star.  We will extend
well-known solutions in two asymptotic limits, the initial thin
expanding shell and the late-time pressure equilibrium between the
ionized gas and its environment, towards an intermediate bridging
regime.  At early time, this requires the extension of the analysis by
\citet{2006ApJ...646..240H} to allow for the thickening of the shocked
shell as its expansion slows towards the sound speed in the neutral
environment; at late time, we generalize the model of oscillations of
an explosively-driven bubble by \citet{1956JAP....27.1152K} to the
boundary conditions appropriate for a photoionized bubble.  A similar
approach was applied by \citet{2009MNRAS.392.1413S} to sound wave
oscillations in galaxy clusters.

\subsection{Thin shell equations}

The early time behaviour of the H{\sc\,ii} region was studied in
detail by \cite{2015MNRAS.453.1324B} using a wide range of numerical
hydrodynamic codes.  That work confirmed that the solution at early
times was well described by the analytic solution of
\cite{2006ApJ...646..240H}.

The motion of the shell is treated using rate of change of the total
momentum of the swept-up shell
\citep{1977ApJ...214..725E,2006ApJ...646..240H}.  Assuming that the
shell is thin, this gives
\begin{equation}
\frac{d}{dt}(M\dot{R}) = 4\pi R^2 (P_1-P_0) \;,\label{e:hhfront}
\end{equation}
which also includes the correction for the effect of the pressure in
the external medium introduced by \citet{2012MNRAS.419L..39R}, and in
which
\begin{align}
M &= \frac{4\pi}{3}\rho_0 R^3\\
P_1 &= \rho_1 \ci^2 \\
P_0 &= \rho_0 c_0^2 \\
\rho_1 &= \rho_0\left(\frac{R_{\rm St}}{R}\right)^{3/2}
\end{align}
are the equation of mass conservation, the isothermal equations of
state assumed in the ionized and neutral material,
and the condition for the ionization equilibrium.  The initial
(constant) background medium has density $\rho_0$ and pressure $P_0$;
and the isothermal sound speeds in neutral ($c_0$) and ionized ($\ci$)
gas are constant parameters of the model, as is the initial
Str\"omgren radius of the photoionized region.

Substituting in, we find an equation equivalent to the system derived
by \citet{2012RMxAA..48..149R}, and referred to as `Raga-II' by
\citet{2015MNRAS.453.1324B},
\begin{equation}
{d\over dt}\left(R^3\dot{R}\right) = 
3R^2\left[\ci^2\left(R_{\rm St}\over R\right)^{3/2} - c_0^2\right].
\label{e:second}
\end{equation}
This equation has been constructed so that a static bubble at the
stagnation radius is an equilibrium solution.  However, it is
symmetric in time, so its solutions must also be symmetric, either
repeating with a finite period or extending to $t\to\pm\infty$.  This
explains the periodic oscillating behaviour of the solutions in the
limit of rapid recombination found by \citet{2012RMxAA..48..149R}.

Multiplying equation~(\ref{e:second}) by $R^3\dot{R}$, it can be
integrated to give
\begin{equation}
{1\over 2}\left(R^3\dot{R}\right)^2 = 
\left[{2\over 3}\ci^2 R_{\rm St}^{3/2} R^{9/2} - {1\over 2}c_0^2R^6\right] 
+{\rm const}.\label{e:first}
\end{equation}

If we assume that both $c_0$ and the integration constant are zero,
then the equation can be integrated again to give the
\cite{2006ApJ...646..240H} solution
\begin{equation}
  R = R_{\rm St}\left(1+{7\over 4}\sqrt{4\over3}{\ci t\over R_{\rm St}}\right)^{4/7},
  \label{e:hiint}
\end{equation}
where we have taken $R=R_{\rm St}$ at $t=0$ to define the constant of
integration (which is simply a time offset of the solution).

Ignoring the term in $c_0$ is reasonable so long as $\dot{R} \gg c_0$.
The assumption that the constant of integration in
equation~(\ref{e:first}) can be taken as zero is more arbitrary, being
a requirement in order for the system to have a simple closed form
solution, rather than being determined from the initial conditions of
the system.  Equation~(\ref{e:hiint}) gives an initial velocity of $2
\ci/\sqrt{3}$ at $R=R_{\rm St}$; equation~(\ref{e:first}) can be
integrated numerically for differing values of this initial velocity,
but the results tend to converge with time.

In Figure~\ref{f:approx_if_shock}, this approximation is compared to
our numerical integral.  At early time, the expansion rate agrees well
\citep[as already demonstrated
  by][]{2012RMxAA..48..149R,2015MNRAS.453.1324B}.  At later time,
however, the shock wave moves away more rapidly than
equation~(\ref{e:hiint}), while the expansion of the ionization front
stalls.  Clearly, a more detailed model for the structure of the
H{\sc\,ii} region is required to derive a more accurate analytic
approximations for the expansion of shock and ionization fronts.

\subsection{Thick shell approximation}

Equation~(\ref{e:hiint}) does not differentiate between the radius of
the ionization front and that of the shock front, i.e.\@ it assumes
that the swept-up shell of material between them is thin.  This is a
reasonable assumption when the outgoing shock speed is far greater
than the sound speed in the neutral material, as the isothermal
equation of state means that the shocked material compresses by a
factor the square of the shock Mach number.  However, as the
H{\sc\,ii} region develops, the region of swept-up neutral material
thickens and these fronts move apart.  To improve the accuracy of the
formulation at later times, this distinction must be taken into
account.

We consider the expansion of the swept-up shell as it becomes
geometrically thick.  Assuming that the material within it has
approximately constant velocity, the rate of change of its radial
momentum can be modelled by generalizing equation~(\ref{e:second}) to
become
\begin{equation}
{d\over dt}\left(M(R_{\rm SF})\dot{R}_{\rm shell}\right) = 
4\pi R_{\rm SF}^2\rho_0\left[\ci^2\left(R_{\rm St}\over R_{\rm IF}\right)^{3/2} - 
  c_0^2\right].\label{e:thick}
\end{equation}
where the terms in $R$ in equation~(\ref{e:second}) have been modified
to distinguish $R_{\rm SF}$ as the radius of the shock, $R_{\rm IF}$
as the radius of the ionization front, and $\dot{R}_{\rm shell}$ as
the velocity of material in the shocked shell.  From the isothermal
jump conditions, we take
\begin{equation}
    \dot{R}_{\rm shell} = \dot{R}_{\rm SF} - {c_0^2\over\dot{R}_{\rm SF}},
\end{equation}
which can be substituted in equation~(\ref{e:thick}) to derive an
second-order equation for $\dot{R}_{\rm SF}$.  This relation between
the shock velocity and shell expansion rate was also used by
\citet{2012MNRAS.419L..39R}.

Equation~(\ref{e:thick}) assumes approximate pressure balance through
the shell.  The outward force on the shell includes both the force at
the ionization front $4\pi R_{\rm IF}^2 P_1$ and the hydraulic
amplification through the shell, i.e.\@ it is
\begin{equation}
    F \simeq 4\pi R_{\rm IF}^2 P_1 +
    \int_{R_{\rm IF}}^{R_{\rm SF}}
    \left[2P(r)\over r\right]4\pi r^2{\rm d}r \simeq 4\pi R_{\rm SF}^2 P_1.
\end{equation}

The shell thickness can be derived from mass conservation and
approximate pressure balance at the ionization front, assuming that
the shell has constant density, so
\begin{equation}
R_{\rm IF}^{3/2} = {2 \ci^2 R_{\rm St}^{3/2}\over
c_0^2+\sqrt{c_0^4+
4\ci^2\left(\ci^2-c_0^2\right)\left(R_{\rm St}/R_{\rm SF}\right)^3}}.\label{e:iffromsf}
\end{equation}
If $R_{\rm SF} = R_{\rm St}$, or $c_0/\ci \to 0$, $R_{\rm IF} =
R_{\rm SF}$ as expected.  If $R_{\rm SF}\to \infty$, $R_{\rm IF} \to
(\ci/c_0)^{4/3} R_{\rm St}$, which is also the late-time equilibrium
solution.  This equation for the ionization front radius in terms
of the shock front radius allows the equation for the shock front
position to be integrated.

From Figures~\ref{f:approx_if_shock} and~\ref{f:approx_if_early}, we
see that the thick shell equations capture the overall behaviour of
the ionization front and shock front quite well, in particular the
eventual expansion of the shock front at the sound speed in the
neutral medium, and the final equilibrium radius of the ionization
front.  However, in the numerical calculations, the inertia of the
shell leads the ionization front to overshoot, an effect which cannot
be captured as a result of the simplified model of the structure of
the swept-up shell.  We will now develop a more detailed model to
capture the overshoot and relaxation of the ionization front radius.

\subsection{Late time behaviour of the H{\sc\,ii} region}

At late time, the leading shock expands to a distance $R_{\rm SF}\gg
R_{\rm Stag} = (\ci/c_0)^{4/3}R_{\rm St}$, so the behaviour of the
H{\sc\,ii} region may be treated independently of the leading shock.
The resulting system is similar to the oscillation of a spherical
bubble in a dense fluid, which was studied by
\cite{1956JAP....27.1152K}.  However, the boundary conditions at the
inner edge of the dense fluid are determined by jump conditions at a
weak D-type ionization front, rather than by a material discontinuity.
In this section, we generalize Keller \& Kolodner's approach to the
late time behaviour of an H{\sc\,ii} region.

The equations governing the motion outside the ionization front are as
given by Keller \& Kolodner, but here we adapt them to the case of
isothermal flow.  The Euler equations for isothermal flow are
\begin{align}
  \dot\rho+\nabla\cdot\left(\rho{\mathbf v}\right) &= 0\\
  \rho \dot{\mathbf v} + \rho{\mathbf v}\cdot\nabla{\mathbf v} &=
  -\nabla p = -c_0^2\nabla\rho,
\end{align}
which in the limit $v\ll c_0$ are solved by a velocity derived from a
velocity potential $\phi$ defined by
\begin{equation}
  {\mathbf v} = \nabla \phi\label{e:vpot}
\end{equation}
which satisfies the isothermal acoustic wave equation
\begin{equation}
\nabla^2\phi = {1\over c_0^2}{\partial^2\phi\over\partial t^2}.\label{e:wave}
\end{equation}

Assuming purely radial flow, as a result of the symmetry of the
problem, we take $\phi = \phi(r,t)$, where the radial velocity is
given by $v = \partial\phi/\partial r$.  Then the momentum equation
becomes
\begin{equation}
  {\partial\over\partial t}\left(\phi'\right) + {1\over
    2}{\partial\over\partial r}\left(\phi'\right)^2 =
    -c_0^2{\partial\over\partial r}\log\rho,
\end{equation}
which may be integrated to give
\begin{equation}
  \dot{\phi} + {1\over 2}\left(\phi'\right)^2 = -c_0^2\log\rho +
  g(t),
\end{equation}
where $g(t)$ is an arbitrary function of $t$ alone.  Since as
$\phi\to{\rm const}$, $\rho\to\rho_0$, the density of the external
medium, we have $g = c_0^2\log\rho_0$.  Hence the density can be
determined from $\phi$ using the isothermal Bernoulli equation
\begin{equation}
-\log\left(\rho\over\rho_0\right) = 
{1\over c_0^2}\left(\dot\phi + {1\over 2}(\phi')^2\right).\label{e:isobernoulli}
\end{equation}

The solution to equation~(\ref{e:wave}) satisfying the boundary
conditions of the problem is a general outgoing wave
\begin{equation}
  \phi = {1\over r} f\left(t-r/c_0\right).\label{e:wavesol}
\end{equation}
Here $f$ is an arbitrary function which is determined by the matching
conditions at the ionization front jump conditions for mass and
momentum flux
\begin{align}
-\rho_i\dot{r}_i &= \rho\left(v-\dot{r}_i\right)\\
\rho_i \left[\dot{r}_i^2 + \ci^2 \right]
&= \rho\left[ \left(v-\dot{r}_i\right)^2 + c_0^2\right],
\end{align}
where we assume that the velocity of the ionized gas in the rest frame
of the system as a whole can be taken as zero.  When the ionization
front is reducing in radius, it will become a recombination front, but
the same jump conditions will apply in the D-type limit.  In the limit
$v\ll c_0$, the boundary conditions at the D-type front at the edge of
the ionized region will be
\begin{align}
    v &= \beta \dot{r}_i\\
    \rho_i \ci^2 &= \rho c_0^2,
\end{align}
where we define
\begin{equation}
\beta = \left(1-{c_0^2\over \ci^2}\right) \simeq 1.
\end{equation}

We also assume the Str\"omgren condition applies with uniform density
in the ionized region
\begin{equation}
  \Qzero = {4\over 3}\pi r_i^3\alpha_{\rm B} {\rho_i^2\over \mu_i^2}
  +4\pi r_i^2{\rho_i\over \mu_i}\dot{r}_i,
\end{equation}
so
\begin{equation}
  {\rho\over\rho_0} = {1\over \xi+\sqrt{1+\xi^2}}\left(R_{\rm
    Stag}\over r_i\right)^{3/2},
\end{equation}
where
\begin{equation}
  \xi = {3\over2}\left(r_i\over R_{\rm Stag}\right)^{1/2} {\ci^2\over
    c_0^2} {\mu_i\over\alpha_{\rm B}}\dot{r}_i
\end{equation}
is typically small.

Given these boundary conditions, together with the definition of the
velocity potential, equation~(\ref{e:vpot}), the Bernoulli
equation~(\ref{e:isobernoulli}) and the wave-like solution,
equation~(\ref{e:wavesol}), we have
\begin{align}
\beta \dot{r}_i + {f'\over r_i c_0} &=  -{f\over r_i^2}, \\
\beta^2 \dot{r}_i^2 + 2{f'\over r_i} &=
3c_0^2\log\left(r_i\over R_{\rm Stag}\right)-
2c_0^2\sinh^{-1}\xi.
\label{e:bernoulli_fr}
\end{align}
In what follows, we will ignore the higher-order term in $\xi$ for
simplicity.  Eliminating $f'$ between these equations gives an
equation for $f$ in terms of $r_i$ and $\dot{r_i}$
\begin{equation}
    {2c_0 f\over r_i^2}+3c_0^2\log\left(r_i\over R_{\rm Stag}\right)
    =\left(\beta\dot{r_i}-2c_0\right)\beta\dot{r_i}.
    \label{e:fsol}
\end{equation}
To eliminate $f$ from this equation, we can differentiate with respect
to $t$, and substitute for $f'$ in the result using
equation~(\ref{e:bernoulli_fr}).  Note that as these expressions are
evaluated at the ionization front,
\begin{equation}
  \left.\partial f\over \partial t\right\vert_{r_i} =
  \left(1-{\dot{r}_i\over c_0}\right) f',
\end{equation}
and that $R_{\rm Stag}$ can vary in time if $Q_0$ is not constant.
The result is a nonlinear second order ordinary differential equation
for $r_i$,
\begin{align}
    {2\over3}r_i\left(c_0-\beta\dot{r_i}\right)\beta\ddot{r_i} 
    + c_0^2\dot{r_i} 
    + c_0^2\left(c_0+\dot{r_i}\right)\log\left(r_i/ R_{\rm Stag}\right)\nonumber \\
    = {1\over3}\beta\dot{r_i}^2\left[\beta(c_0+\dot{r_i})-4c_0\right]
    +{1\over 3}c_0^2 r_i {\dot{Q_0} \over Q_0}.\label{e:nonlin}
\end{align}
When this is solved, equation~(\ref{e:fsol}) can be used to determine
$f$, and hence the full solution for $\rho$ and $v$ between the
ionization front and the leading shock.

We will first look for the linear solution to the system with a
constant ionizing flux, in the limit $c_0 \ll \ci$, which will capture
the asymptotic behaviour of the front as it relaxes to its equilibrium
stagnation radius.  Writing $r_i = R_{\rm Stag}+\delta$, using Taylor
expansions for power and logarithmic terms, and ignoring terms of
higher than linear order in perturbations from equilibrium, we find
that the offset of the ionization front from its equilibrium position
satisfies
\begin{equation}
    {2\over 3} R_{\rm Stag}^2\ddot{\delta}
    + c_0 R_{\rm Stag}\dot\delta + c_0^2\delta = 0,\label{e:linear}
\end{equation}
and hence
\begin{equation}
\delta = \delta_0 \cos(\omega t+\psi)\exp(-\lambda t) ,
\end{equation}
with $\omega = (15/16)^{1/2} c_0/R_{\rm Stag}$, 
$\lambda = (3/4)c_0/R_{\rm Stag}$, and where the amplitude $\delta_0$ and phase $\psi$ are arbitrary constants.

As $\beta$ is close to $1$, the major effect causing the rapid damping
of oscillations for an H{\sc\,ii} region seen in
Figure~\ref{f:approx_if_late}, compared to an isothermal bubble with a
constant mass of hot gas \citep{1956JAP....27.1152K}, must be the
different manner in which the pressure within the H{\sc\,ii} region
varies with radius, rather than the mass flux through its surface.

As seen in Figure~\ref{f:approx_if_late}, the linear asymptotic
solution gives an excellent fit to the relaxation of the ionization
front at late time found in the numerical calculation, but
Figure~\ref{f:approx_if_early} shows that it diverges at earlier
times.  We will now investigate what modifications can be made to the
system to improve the fit to the ionization front radius at all times.

A first approach would be to include the higher order terms in
equation~(\ref{e:nonlin}) which were ignored in deriving
equation~(\ref{e:linear}).  However, equation~(\ref{e:nonlin}) has a
singular point, where the factor multiplying the highest derivative
becomes zero, when $\beta\dot{r_i} = c_0$.  Continuous solutions
passing through this speed are possible only at the ionization front
radius
\begin{equation}
  r_i = R_{\rm Stag}\exp\left(-{6-\beta\over 3(\beta+1)}\right)
  \simeq 0.368R_{\rm Stag}.
\end{equation}
However, in numerical simulations, the outward velocity of the front
is substantially larger than $c_0$ when it reaches this radius,
suggesting that the loss of accuracy of the approximations used when
deriving equation~(\ref{e:nonlin}) is a more immediate limitation to
the accuracy of the solution than the presence of the critical point.

\begin{figure}
\begin{center}
\includegraphics[width=\linewidth]{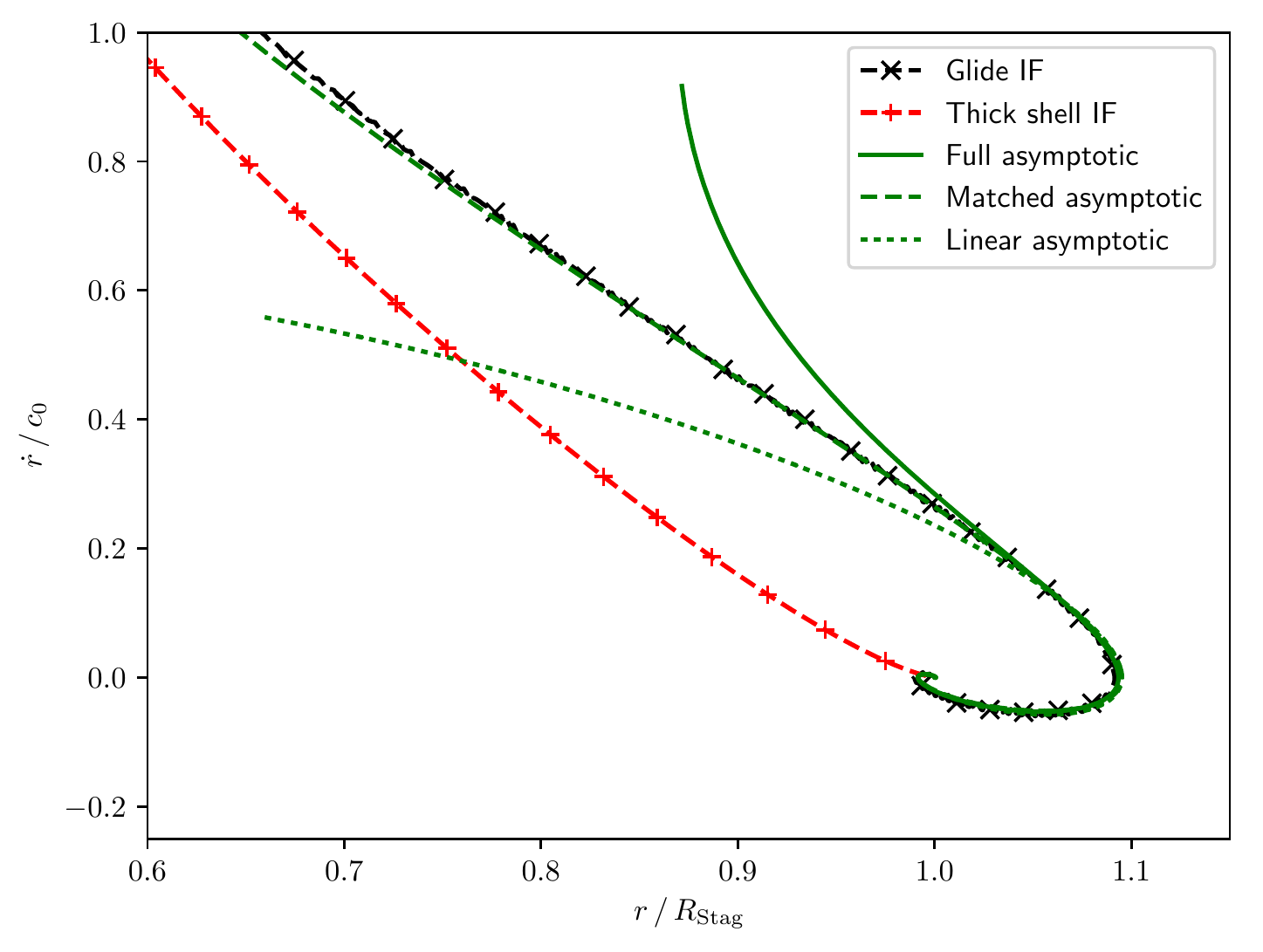}
\end{center}
\caption{As Figure~\protect\ref{f:approx_if_shock}, but showing front
  velocity as a function of radius.}
\label{f:approx_if_phase}
\end{figure}

To the order of accuracy of the derivation leading up to
equation~(\ref{e:nonlin}), we can approximate
\begin{equation}
1-\beta \dot{r_i}/c_0 \simeq {1\over 1+\beta \dot{r_i}/c_0},
\end{equation}
which removes the singularity.  Figure~\ref{f:approx_if_late}
includes a comparison with the numerical solution of
equation~(\ref{e:nonlin}) with this modification to the leading term
(labelled as the `Full asymptotic' case).  While including additional
terms in the approximation makes the semi-analytic solution more
accurate at intermediate times, it diverges from the results of the
full dynamical simulation rapidly at earlier times.  This behaviour is
fairly typical when a higher order solution is based on expansions at
a single time, where increasing accuracy around this point can lead to
more rapidly divergent behaviour away from it.  As we will see next, a
better representation may be derived by considering the asymptotic
behaviour of the solution both at early and late times.

As equation~(\ref{e:nonlin}) is a second-order autonomous ordinary
differential equation, it may be converted into a system of two
independent first-order ODEs, one of which is formally independent of
time.  Taking $q=\dot{r_i}/c_0$, the second time derivative may be
replaced by $\ddot{r_i} = c_0^2q {\rm d}q/{\rm d}r_i$, to become
\begin{align}
    {2\over3}\left(1-\beta q\right)\beta q{{\rm d}q\over {\rm d} r_i}
    + q + \left(1+q\right)\log(r_i/R_{\rm  Stag})
     \nonumber \\
    ={1\over3}\beta q^2\left[\beta q-(4-\beta)\right].\label{e:nonlinq}
\end{align}
Plotting $q=\dot{r_i}/c_0$ against $r_i$ thus allows the results of the
various approximations to be compared independent of any arbitrary
time offset.  Figure~\ref{f:approx_if_phase} shows the velocity of the
ionization front as a function of its radius, derived from the
simulated data and for various approximations.  The linear
approximation to the late time behaviour forms a logarithmic spiral,
while the full asymptotic solution diverges from the numerical results
in the opposite direction.

\subsection{Matched asymptotic solution}

We will now look for an equation for the position of the ionization
front which can capture its behaviour throughout its evolution with
good accuracy, by modifying equation~(\ref{e:nonlin}) with terms which
are small at late time, but which will mean that it agrees with the
thin shell system, described by equation~(\ref{e:second}), at early
times.

Equation~(\ref{e:second}) can be expressed as
\begin{equation}
  R\ddot{R} + 3\dot{R}^2 = 3
  c_0^2\left[\left(R_{\rm Stag}\over R\right)^{3/2} - 1\right],
  \label{e:secondmod}
\end{equation}
and we will now look to match terms at the same order
equation~(\ref{e:nonlin}) while we can still assume $R\simeq r_i$.
Adding lower-order terms to equation~(\ref{e:secondmod}) so that it
matches the behaviour of the solution at late time, term-by-term in
Taylor expansions, in particular requiring that the linear-order terms
which control the late-time behaviour in Equation~(\ref{e:nonlin}) are
exact, gives
\begin{align}
  \beta r_i \ddot{r}_i &+{1\over 2}\beta(7-\beta) \dot{r}_i^2 + {3\over
    2}{\dot{r}_ic_0\over 1+C\dot{r}_i^2/c_0^2}
  +\nonumber\\ &={c_0+3B\dot{r}_i\over
    c_0+B\dot{r}_i}c_0^2\left[\left(R_{\rm Stag}\over
    r_i\right)^{3/2}-1\right]
  +{1\over 2} c_0 r_i {\dot{Q}_0\over Q_0},\label{e:matched}
\end{align}
where $B=(1+\beta)/2$ and $C=2.25$, and we have included the source
term required if the luminosity of the source varies.  This modified
form accounts for both the effects of the finite temperature of the
neutral material, and the late time relaxation to equilibrium.  The
constant $B$ is determined by the matching procedure, but the
reduction of the damping term linear in $\dot{r}_i$ controlled by the
dimensionless parameter $C$ is not.  This reduction in the linear
damping term is found to be necessary to obtain a good fit to the full
numerical evolution, even though the quadratic term is larger at early
time, when the expansion velocity of the ionization front $\dot{r}_i >
c_0$.

As can be seen in Figure~\ref{f:approx_if_late}, the solution of this
matched asymptotic system for the ionization front position,
equation~(\ref{e:matched}) fits the data from the numerical solution
with good accuracy throughout.  A small phase-shift is apparent in the
behaviour at late time compared to the linear solution, as the phase
is no longer a free parameter which can be varied to optimize the
agreement.  Figure~\ref{f:approx_if_phase} shows excellent agreement,
showing that this phase-shift corresponds to a small time offset in
the late-time behaviour.

\begin{figure}
\begin{center}
\includegraphics[width=\linewidth]{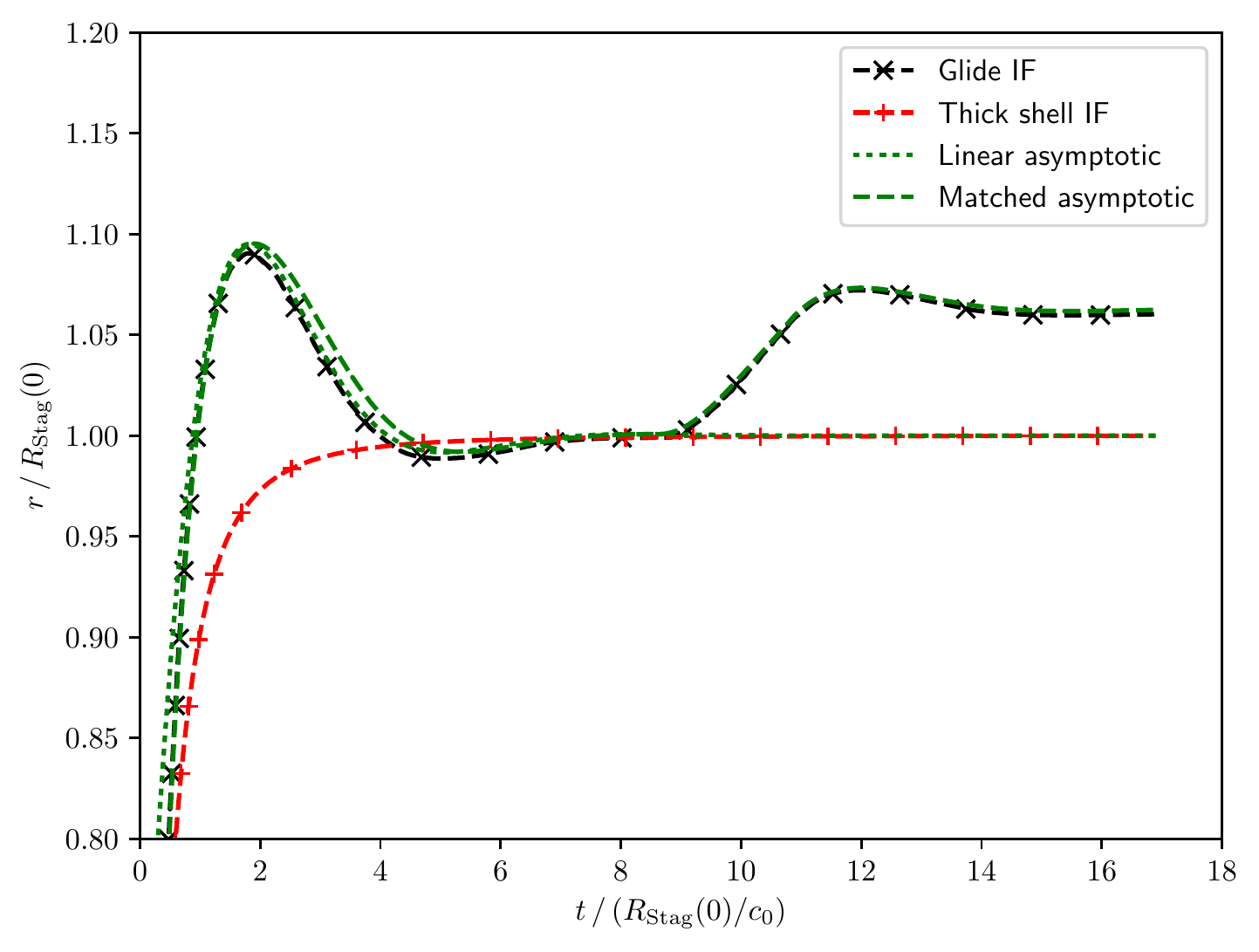}
\end{center}
\caption{As Figure~\protect\ref{f:approx_if_shock}, but scaled to
  emphasize late-time relaxation, with ionizing luminosity linearly
  increasing to 120 per cent of its initial value between $100$ and
  $125{\rm\,Myr}$ (i.e.\@ $t c_0/R_{\rm Stag}(0) = 8.6$ and $10.8$).
  Note that the time and radius are scaled to the initial stagnation
  radius, $R_{\rm Stag}(0)$. The change in luminosity is not
  implemented in the linear asymptotic form or thick shell forms,
  which are included simply for cross-reference.}
\label{f:approx_variable}
\end{figure}
Equation~(\ref{e:matched}) is general enough to allow for the
variation of the ionizing luminosity of the source with time, by
varying $R_{\rm Stag}$ to reflect the changing value of $\Qzero$.  In
Figure~\ref{f:approx_variable}, we compare the results of
equation~(\ref{e:matched}) with the variation of ionization front
radius when the ionizing luminosity is assumed to increase linearly to
$1.2\times 10^{49}$ from $100{\rm\,Myr} = 8.64 R_{\rm Stag}/c_0$ to
$125{\rm\,Myr}$.  The agreement between the model and the numerical
solution as the ionizing flux varies is excellent, as expected so long
as the relative rate of change of ionizing flux is slow compared to
the sound crossing time of the ionized bubble.

\section{Discussion and conclusions}

In this paper, we have extended the analytic study of the expansion of
an H{\sc\,ii} region in the idealized case of a constant source and
uniform ambient medium. The agreement between model and simulation
during the D-type expansion phase has been improved compared to
previous works, to within 2 per cent relative error at all times.  A
thick-shell approximation allows for the stalling of the ionization
front and decay of the leading shock to a weak compression wave as it
escapes to large radii.  An acoustic approximation, modifying the work
of \citet{1956JAP....27.1152K} to take account of the different
boundary conditions at the ionization front, captures the late-time
oscillations of the H{\sc\,ii} region about the stagnation radius, and
its response to variations in ionizing luminosity.

We have examined the dynamics of H{\sc\,ii} regions in significantly
greater depth than previous work; the thick-shell solution and the
asymptotic solutions give considerable physical insight that allow us
to understand all features of the numerical solutions.  This work also
resolves all of the small disagreements that we are aware of in the
literature comparing numerical simulations with analytic solutions,
once again showing the importance of having benchmark analytic
solutions to test problems for validation of simulation codes.  The
dynamical response of the system on the crossing time of the ionized
bubble at the sound speed of neutral material, and the rapid decay in
intensity of radial oscillations, are likely to be general features of
the response of H{\sc\,ii} regions to external and internal
perturbations.

The mathematics and physics of spherically expanding H{\sc\,ii}
regions turns out to be significantly richer and more complex than
initially thought before the \textsc{StarBench} project
\citep{2015MNRAS.453.1324B}.  We anticipate that the inclusion of
multi-dimensional effects such as dynamical instability will be even
more interesting, even using the dramatically simplified thermal
physics considered here.

\section{Acknowledgements}
The authors thank contributors to the \textsc{StarBench} project for
useful discussions.  The authors are grateful to the Dublin Institute
for Advanced Studies, and to Hilary O'Donnell for her hospitality and
assistance, for facilitating a short residential workshop at Dunsink
Observatory in May 2016 during which this project was significantly
advanced.

TJH is funded by an Imperial College Junior Research Fellowship.
JM acknowledges funding from a Royal Society--Science Foundation Ireland University Research Fellowship.

This paper contains material \copyright{} British Crown Owned
Copyright 2017/AWE.

\bibliographystyle{mnras}
\bibliography{refs}

% Don't change these lines
\bsp	% typesetting comment
\label{lastpage}
\end{document}